\documentclass[letterpaper, 10 pt, conference]{ieeeconf} 
\IEEEoverridecommandlockouts                              % This command is only
\usepackage{multirow}
\usepackage{float}
\floatstyle{plain}
\newfloat{bottomfigure}{bp}{lop}
\usepackage{cite}
\usepackage[cmex10]{amsmath}
\usepackage{mathtools}
\usepackage{amssymb}
\usepackage{amsmath}
\usepackage{color}
\usepackage[caption=true]{caption}
\usepackage[font=footnotesize]{subfig}
  \usepackage{epsfig}

\def\b0{\mbox{{\bf 0}}}

\def\bA{\mbox{{\bf A}}}
\def\bB{\mbox{{\bf B}}}

\def\bD{\mbox{{\bf D}}}

\def\bH{\mbox{{\bf H}}}
\def\bI{\mbox{{\bf I}}}

\def\bK{\mbox{{\bf K}}}
\def\bL{\mbox{{\bf L}}}

\def\bR{\mbox{{\bf R}}}

\def\bU{\mbox{{\bf U}}}

\begin{document}
\title{Stability Analysis of Network of Similar-Plants via Feedback Network}
\author{Behzad Shahrasbi\\
School of Electrical Engineering and Computer Science\\
University of Central Florida, Orlando, FL 32816 \\
Emails: behzad.shahrasbi@knights.ucf.edu}
\maketitle
\date{\today}

\begin{abstract}
Here, we study a networked control system with similar linear time-invariant plants. Using master stability function method we propose a network optimization method to minimize the feedback network order in the sense Frobenius norm.  Then we verify our results with numerical example. We show that this method outperforms the known feedback network optimization methods namely matching condition.

 \end{abstract}
%%%%%%%%%%%%%%%%%%%%%%%
%%%%%%
%%%%%%%%%%%%%%%%%%%%%%%
\section{Introduction}
Recently the study of interacting dynamical systems as a network have drawn the attention of most scientists and engineers in various fields of physics, biology, and such \cite{Winfree67,Manaffam14,SyncNet,NavigationSmall,Manaffam13a}. The main stream of these efforts have been focused on stability of the dynamical flow in the network \cite{SyncNet,MasterEq}. The other aspect of the efforts has been concentrated on the separation of structural properties of the network from local dynamics of the network \cite{Manaffam13a,weightedNet,Networks,Heterogeneity1,NatureSmall,Boundcomplex}.\

One of the important categories in the subject of dynamical systems is designing a network of controllers based on available inputs from either the local information or in general sense from the other parts of the network to asymptotically stabilize the entire network. The first approach, in which the local state information is used to form the control law for each subsystems, is called decentralized control. This approach usually is used in network of sparsely or weakly coupled systems, since the the dynamical flow in the richly connected networks cannot always be met up by the solely local information \cite{trig2010}.\

The other approach uses the state information of the other nodes as well as local state data and it is known as networked control system. Networked control systems can be used in the wide range of the observable networks to reach the asymptotic stable state \cite{NCS2011,Manaffam12,Manaffam13b,Manaffam13c,Manaffam15,trig2010}. Since this method uses the information from wider subset of the network, with smart enough design it can assure the stability of a controllable network, more efficiently\cite{MasterEq,Manaffam13a,Manaffam14}. However, this method  attains the goal by increasing the complexity of the network by requiring the state estimators at each node and the information subnetwork. The information or feedback network creates the cost of wiring or some other means of communications. Dealing with the problems of estimation errors and uncertainties of communication network are some of important challenges faced by this approach \cite{Manaffam13a,Manaffam15}. \

Here first we try to separate the impact of network structures, such as degree of nodes, minimum and maximum degrees, on the stability from the local dynamics of nodes. Since some of the networks in real world depict the behavior reflected by random networks \cite{NatureSmall,Manaffam13a}, we focus on analyzing the network of plants with random connections in either feedback network or plant network. Assuming a random feedback network which is also a subnet of plant network, we calculate the probability of stability. Then, we extent the scenario to a random network of plants modeled as Erd\"{o}s-R\'{e}nyi network, and compute the probability of stability in extreme cases such as very large network size, weak coupling and etc. \

The rest of the paper is as follow, in following section the system model and required preliminaries to analyze the system is given. Then in the section III we analyze the system, and in the section IV numerical results are presented. And at the end we conclude the study.
\begin{figure}[t]
\begin{center}
\includegraphics[width=3.5in]{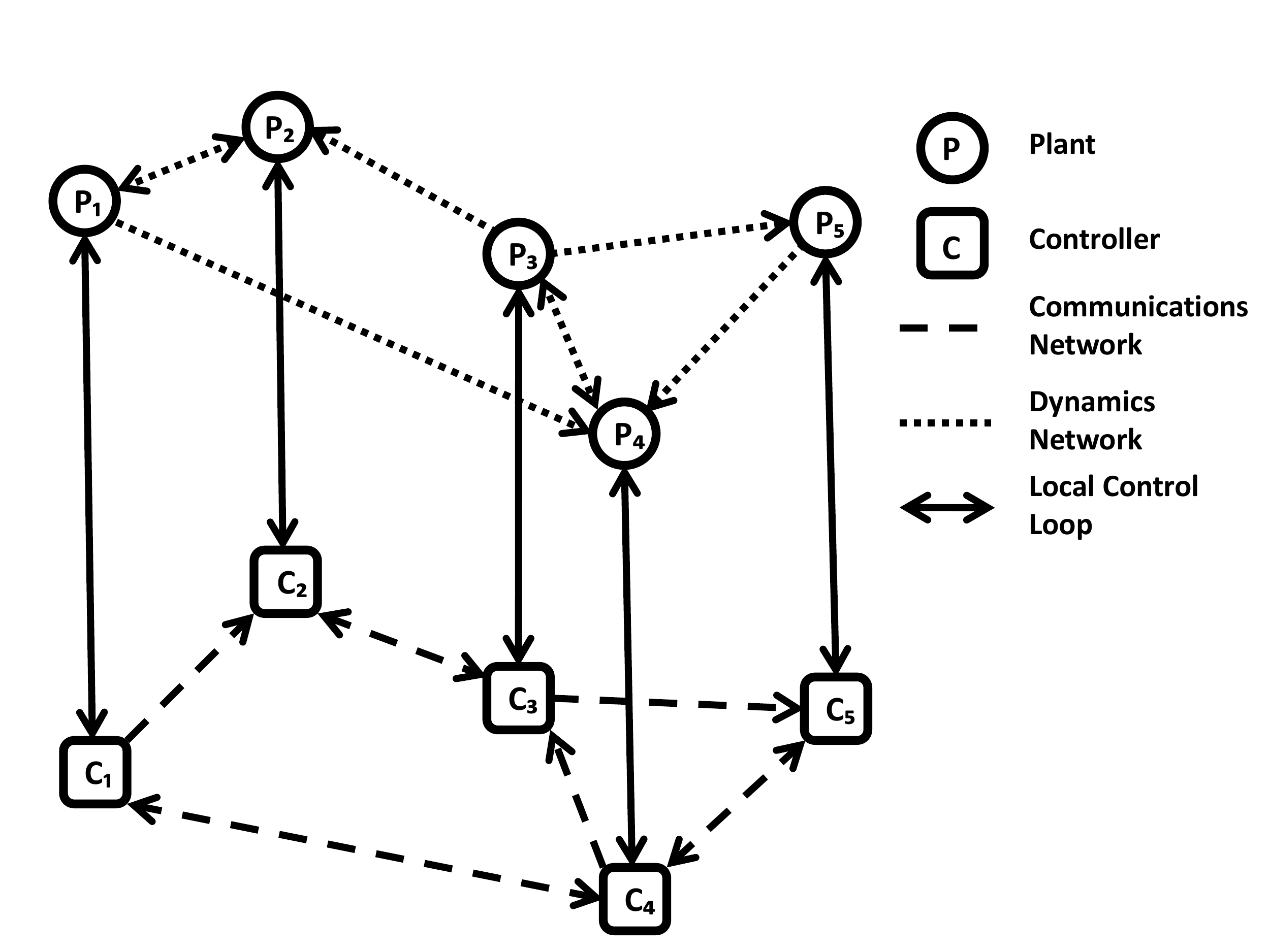}
\caption{Considered Configurations for Networked Control System \cite{Manaffam13b}.}\vspace{-0.7cm}
\label{Fig: Configuration}
\end{center}
\end{figure}
%%%%%%%%%%%%%%%%%%%%%%%
%%%%%%
%%%%%%%%%%%%%%%%%%%%%%%
\section{Notations and System Description}
\subsection{Notations}
The set of real $n$-vectors is denoted by $\mathbb{R}^{n}$ and the set of real $m\times n$ matrices is denoted by $\mathbb{R}^{m\times n}$. Matrices and vectors are denoted by capital and lower-case bold letters, respectively. The Euclidean ($\mathcal{L}_{2}$) vector norm is represented by $||\cdot||$. When applied to a matrix $||\cdot||$ denotes the $\mathcal{L}_{2}$ induced matrix norm, $||\mathbf{A}||=\lambda_{\max}(\mathbf{A}^{T}\mathbf{A})$. Also, we denote a graph corresponding to a network with $\mathcal{G}$, and vector $\mathbf{d}$ denotes the corresponding degree sequence of the graph. 

Table I summarizes the variables used frequently in the rest of the paper.
\begin{table}[t]
\label{vars}
\caption{Frequently used variables}
\begin{tabular}{cl} \hline
{\bf Variable} & {\bf Description}\\ \hline  \hline
$\mathbf{x}_{i}$& State vector of node $i$\\ \hline
 $\mathbf{u}_{i}$& Input vector of node $i$\\ \hline
$\mathbf{D}$& Plant dynamics matrix\\ \hline
$\mathbf{H}$& Coupling matrix between two nodes in the dynamic\\ 
 & network\\ \hline
$\mathbf{R}$& Plant input matrix\\ \hline
$\mathbf{K}$& Local feedback gain\\ \hline
$\mathbf{L}$& Feedback loop gain\\ \hline
$c$& Coupling strength in dynamics network\\ \hline
$N$& Plant network size\\ \hline
\end{tabular}
\end{table}

%%%%%%%%%%%%%%%%%%%%%%%
%%%%%%
%%%%%%%%%%%%%%%%%%%%%%%
\subsection{System Description}
Let assume that we have a network of plants with local dynamical matrix of $\mathbf{D}$ which are coupled by $\mathbf{H}$
\begin{align}
\dot{\mathbf{x}}_{i}=\mathbf{D}\mathbf{x}_{i}+\mathbf{R}\mathbf{u}_{i}+\sum_{j=1}^{N}b_{ji}\mathbf{H}\mathbf{x}_{j},
\end{align}
where $\mathbf{R}$ is the matrix associated with input vector of node $i$, $\mathbf{u}_{i}$ is the local feedback matrix  and $b_{ij}$ is binary variable which denotes the existence of coupling from node $j$ to node $i$. Fig. \ref{Fig: Configuration} shows the configuration of such a network.\\
Now consider the feedback network as
\begin{align}
\mathcal{L}_{i}(\mathbf{x}_{1},~ \cdots,~\mathbf{x}_{N})=\sum_{j=1}^{N}c_{ji}\mathbf{R}\mathbf{L}_{ji}\mathbf{x}_{j},
\end{align}
where $c_{ji}$ is binary variable denoting a feedback from node $j$ to node $i$, $\mathbf{L}_{j}$ is the outgoing feedback from node $j$ to node $i$,. Since we are considering similar plants, $\mathbf{L_{ii}}=K$ and $\mathbf{L}_{ji}=\mathbf{L}$, where $j\neq i$. \\
Thus, the state equation of node $i$ with associated feedback network, $\mathcal{L}_{i}$ is
\begin{align}
\dot{\mathbf{x}}_{i}=(\mathbf{D}+\mathbf{R}\mathbf{K})\mathbf{x}_{i}+\sum_{j=1}^{N}b_{ji}\mathbf{H}\mathbf{x}_{j}+\sum_{j=1}^{N}a_{ji}\mathbf{R}\mathbf{L}\mathbf{x}_{j},
\end{align}
where $a_{ji}=c_{ji}$ if $j\neq i$ and zero otherwise.\\
Defining $\mathbf{F}=\mathbf{D}+\mathbf{R}\mathbf{K}$ and $\mathbf{G}=\mathbf{R}\mathbf{L}$, we have the network state equation as
\begin{align}
\mathbf{x}=(\underbrace{\mathbf{I}_{N}\otimes \mathbf{F}+\mathbf{B}\otimes \mathbf{H}+\mathbf{A}\otimes \mathbf{G}}_{\tilde{\mathbf{F}}})\mathbf{x},
\end{align}
where $\mathbf{I}_{N}$ is identity matrix, $\mathbf{A}=[a_{ij}]$, $\mathbf{B}=[b_{ij}]$, and $\mathbf{x}=[\mathbf{x}_{1}^{T} ~\cdots~\mathbf{x}_{N}^{T}]^{T}$ where the superscript denotes the transpose operator.
%%%%%%%%%%%%%%%%%%%%%%%
%%%%%%
%%%%%%%%%%%%%%%%%%%%%%%
%%----------------------CHAP 3  SYSTEM ANALYSIS -----------
\section{Network design}
The network is asymptotically stable iff the eigenvalues of $\tilde{\mathbf{F}}$ have negative real part. Since \bB~ is a feedback network to be designed, we can alway assume that $\bA$ and \bB~are jointly upper-triangularizable with unitary matrix, \bU, then the eigenvalues of $\tilde{\mathbf{F}}$ are the union of eigenvalues of the matrices of the form $\mathbf{F}+\lambda_{i}\mathbf{H}+\mu_{i}\mathbf{G}$, where $\mu_{i}$ and $\lambda_{i}$ are $i$th eigenvalues of $\mathbf{A}$ and $\mathbf{B}$ under the unitary transformation \bU, respectively (see the Appendix I). If we define the master stability function, $\sigma_{\tilde{\mathbf{F}}}(\lambda,\mu)$, as the maximum real part of eigenvalues of $\tilde{\mathbf{F}}$, then, the network is stable iff  $\sigma_{\tilde{\mathbf{F}}}(\lambda,\mu)<0$.\\
Here, we want to design a minimal feedback network in the sense of squared of Frobenius norm, $\min ||\mathbf{A}||^{2}_{F}=\sum_{i,j}|a_{ij}|^{2}$. 

%%----------------------SUBSECTION Unweighted NETWORK DESIGN-----------

\subsection{Unweighted Feedback Network}
To design a binary network with minimum links, which means choosing a network with minimum degree, we should solve following binary problem
\begin{align}
&\min ||\mathbf{A}||_{F}^{2}=\min \sum_{ij}a_{ij}\\
&\mbox{subject to: }\nonumber\\
&a_{ij} \mbox{'s are binary}\nonumber\\
f_{l i}(\lambda_{i},\mathbf{F},\mathbf{H},\mathbf{G})&\leq\mu_{i}\leq f_{ui}(\lambda_{i},\mathbf{F},\mathbf{H},\mathbf{G}),i\in\{1,~\cdots,~N\}\nonumber
\end{align}
where the interval $[f_{li},~f_{ui}]$ is the closest interval to origin in which $i$th master stability function is negative.\\ This binary optimization problem can be tackled by using Branch and Bound method, which is proven to be $NP$-hard.
%%%%%%% SUBSECTION: Weighted Feedback Network
\subsection{Weighted Feedback Network}
If there is no constraint on the adjacency matrix being unweighted, the problem to solve is reduced to choosing triangular matrix $\mathbf{T}$ such that  $\mathbf{A}=\mathbf{Q}\mathbf{T}\mathbf{Q}^{*}$, and $\mathbf{T}$ has a $[\mu_{1}, ~\cdots,~\mu_{N}=0]$ as its diagonal entries. Thus, $\mathbf{T}$ can be obtained from
\begin{align}
&\min \sum_{ij}||a_{ij}||^{2}\label{eq: minimization1}\\
&\mbox{subject to: }\nonumber\\
f_{l i}(\lambda_{i},\mathbf{F},\mathbf{H},\mathbf{G})&\leq\mu_{i}\leq f_{ui}(\lambda_{i},\mathbf{F},\mathbf{H},\mathbf{G}),i\in\{1,~\cdots,~N\}.\nonumber
\end{align}
If (\ref{eq: minimization1}) has a solution then the resultant weighted feedback network satisfies the master stability conditions, hence, it stabilizes $\mathcal{G}_{P}$.\\
Since the objective function in (\ref{eq: minimization1}) is Frobenius norm-squared of the adjacency matrix of the feedback network, hence we have the following equality
\begin{align}
||A||_F^2=\sum_{ij}||a_{ij}||^{2}=tr(\mathbf{A}^*\mathbf{A})&=tr((\mathbf{QT^*Q^*})(\mathbf{QTQ^*}))\nonumber\\
&=tr(\mathbf{Q(T^*T)Q^*})\nonumber\\
&=tr(\mathbf{T^*T})
\end{align}
where $tr(\cdot)$ denotes the trace operator, and last equality follows from the fact that $\mathbf{Q}$ is unitary. Therefore, we can rewrite (\ref{eq: minimization1}) as
\begin{align}
&\min ||\mathbf{T}||_F^2\label{eq: minimization2}\\
&\mbox{subject to: }\nonumber\\
f_{l i}(\lambda_{i},\mathbf{F},\mathbf{H},\mathbf{G})&\leq\mu_{i}\leq f_{ui}(\lambda_{i},\mathbf{F},\mathbf{H},\mathbf{G}),i\in\{1,~\cdots,~N\}.\nonumber
\end{align}
We know from the Schur decomposition that $\mu_i$'s are the diagonal entries of $\mathbf{T}$. Thus, $||\mathbf{T}||_F^2=\sum_{ij}||t_{ij}||^{2}=\sum_i||\mu_i||^2+\sum_{i,j\neq i}||t_{ij}||^2$. The obvious choice for off-diagonal entries are $t_{ij}=0~j\neq i$, hence, the resultant $\mathbf{T}$ in (\ref{eq: minimization2}) is a diagonal matrix and the problem is reduced to minimization problem for each $i\in\{1,~\cdots,~N\}$ as
\begin{align}
\min ||\mu_i||& \label{eq: minimization3}\\
\mbox{subject to: }&f_{l i}(\lambda_{i},\mathbf{F},\mathbf{H},\mathbf{G})\leq\mu_{i}\leq f_{ui}(\lambda_{i},\mathbf{F},\mathbf{H},\mathbf{G}).\nonumber
\end{align}
This problem has complexity growth linearly with network size $N$.\

Although, it is shown that off-diagonal entries of $\mathbf{T}$ increases the number/weight of links in communication network, in general, those terms can be used to provide diversity which in turns results in more robust control network to avoid congestion, failing links or dealing with other security issues of the network.
%%%%%%%%%%%%%%%%%%%%%%%
%%%%%%
%%%%%%%%%%%%%%%%%%%%%%%
%%----------------------CHAP 5  NUMERICAL RESULTS-----------

\section{Numerical Results}
For the purpose of numerical demonstration, let 
\begin{align*}
	\bD=\begin{array}{cc}\left[ \begin{array}{cc}
		3 & 5\\
		-1 & 0
		\end{array}
		\right] , 	
		&
		 \bR=\left[ \begin{array}{c}
		1\\
		0
		\end{array}
		\right] ,
		\end{array}
\end{align*}
and
\[\bH=\left[ \begin{array}{cc}
		1 & 0\\
		0 & 0
		\end{array}
		\right]  .
\]
Lets choose 
\begin{align*}
\bK= -\left[ \begin{array}{cc}
		5 & 0
		\end{array}
		\right] 
\end{align*}
		and to satisfy the matching condition (see for example \cite{trig2010}) $\forall i,j$
		\begin{align*} 
\bL_{ji}=-\left[ \begin{array}{cc}
		1 & 0
		\end{array}
		\right] .
\end{align*}
\begin{figure}[t]
\begin{center}
\includegraphics[width=3.2in]{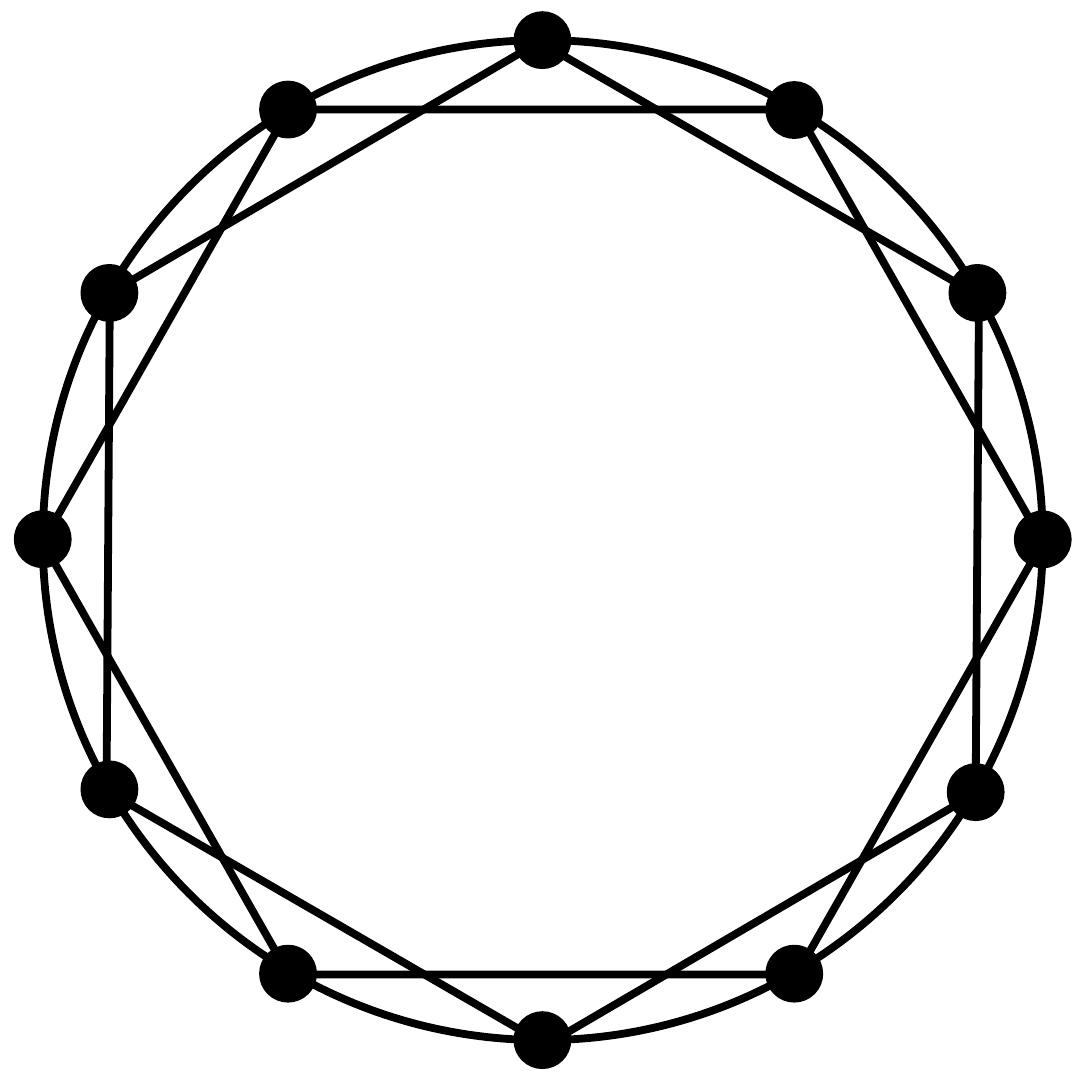}
\caption{4-regular ring network.}\vspace{-0.7cm}
\label{fig: NormvsN}
\end{center}
\end{figure}
\begin{figure}[t]
\begin{center}
\includegraphics[width=3.7in,height=3.5in]{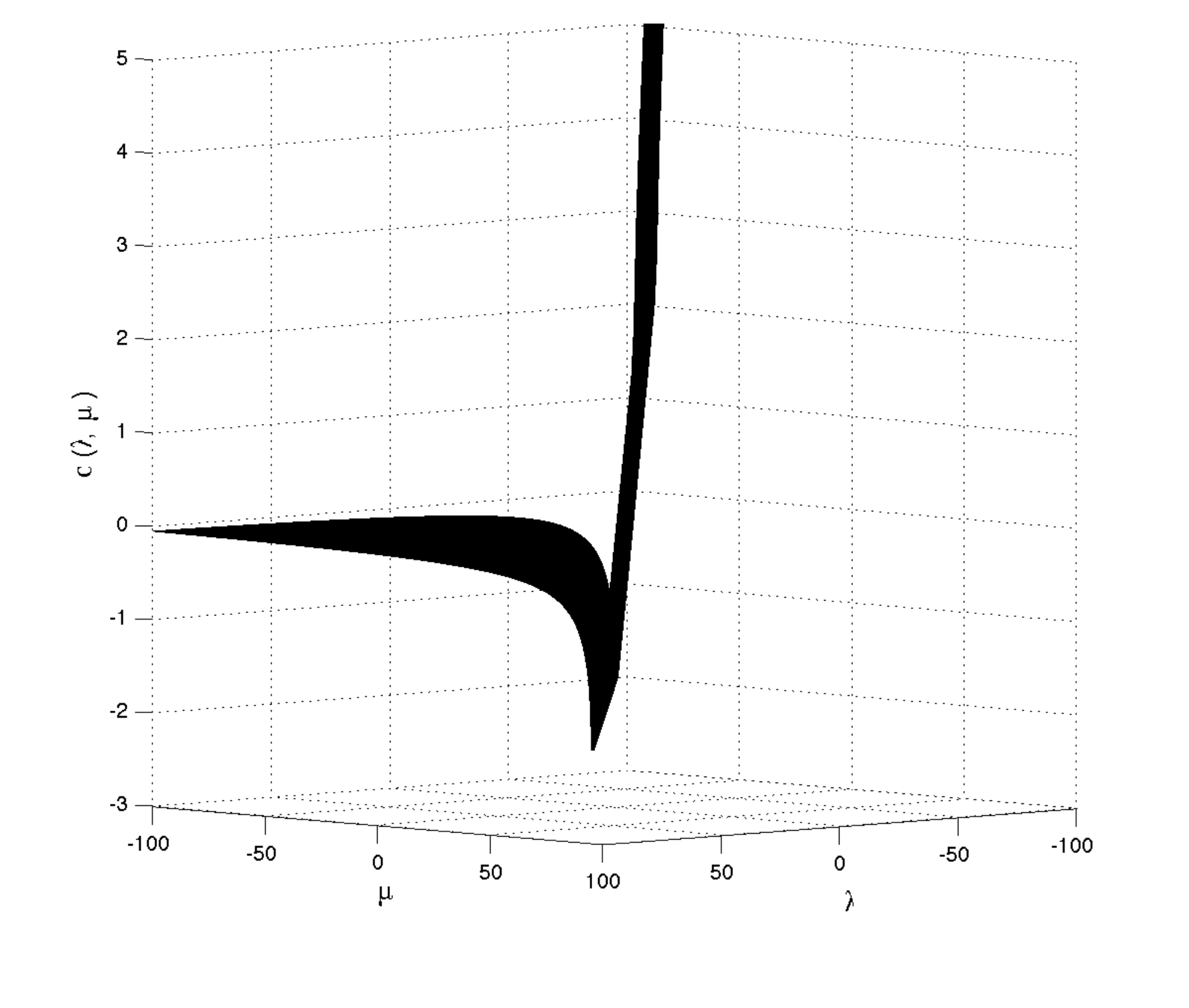}
\caption{Master stability function of the network as a function of $\lambda$ and $\mu$.}\vspace{-0.7cm}
\label{fig: MSF}
\end{center}
\end{figure}
Then we will have the master stability function given in Fig. \ref{fig: MSF} as a function of $(\lambda,~\mu)$, where $\lambda$ and $\mu$ represent the eigenvalues of the plant network and the eigenvalues of feedback network, respectively. The region which returns negative value for master stability function is the stable region and all the possible feedback networks to make the overall network stable falls into this region.\
\begin{figure}[t]
\begin{center}
\includegraphics[width=3.7in,height=3.7in]{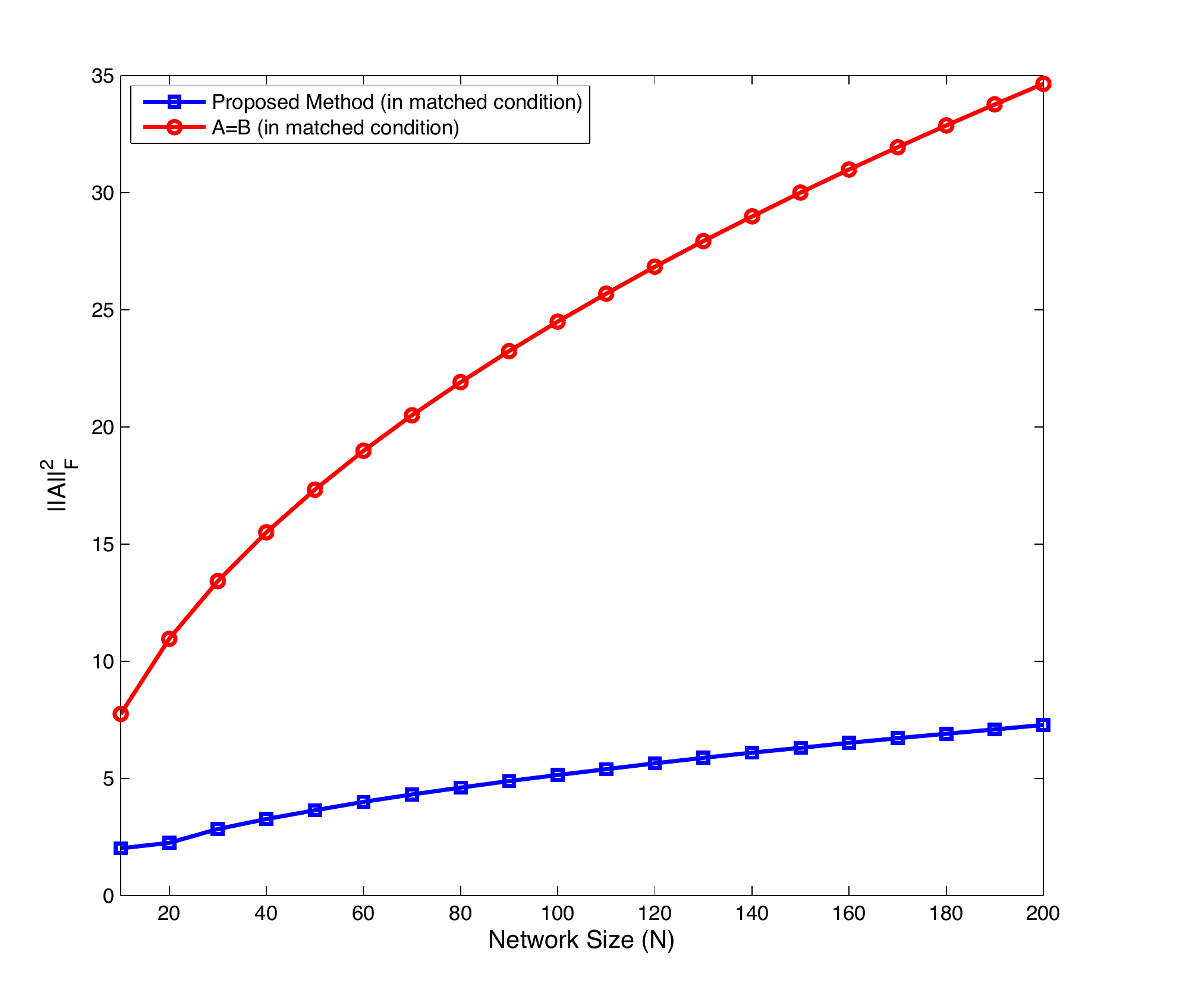}
\caption{Comparison of Frobenius norm of the connections between the proposed method (weighted one) and A=B.}\vspace{-0.7cm}
\label{fig: NormvsN}
\end{center}
\end{figure}

Fig. \ref{fig: NormvsN} shows the Frobenius norm of communication/feedback network versus network for both matching condition proposed in \cite{trig2010} and our proposed method for a ring plant network with $k=4$-degree (see Fig. 2). As it an be seen the feedback network size for our method is substantially lower than that if \cite{trig2010} under matching condition, i.e., $\|\bR\bL-\bH\|=0$.

Now consider the globally connected plant network of size $N=8$
\begin{align}
\bA = \left[\begin{array}{ccccc}
		0 & 1 &1 & \cdots & 1\\
		1 & 0 &1 & \cdots& 1\\
		1 &1& 0 &\ddots &\vdots\\
		\vdots& \vdots&\vdots&\ddots&\vdots\\
		1& 1 & 1 &\cdots &0
		\end{array}
		\right]
\end{align}
the corresponding optimal feedback network in the sense of Frobenius norm is 
\[\bB= \frac{62}{99} \bA\]
with $\|\bB\|_F = \frac{62}{99}N= \frac {496}{99}\approx 5.01$ compared to matching method \cite{trig2010} which is $\|\bB\|_F = \sqrt{N(N-1)}=\sqrt{56}\approx 7.48$.
		 
\section{Conclusion}
Here we proposed a network optimization method based on master stability function to minimize the feedback network order in the sense Frobenius norm. As our results have shown our method always outperforms the benchmark method of matching condition. As a future research topic it is interesting to see if this method can be generalized to problem of synchronization of nonlinear oscillators.
%%%%%%%%%%%%%%%%%%%%%%%
%%%%%%
%%%%%%%%%%%%%%%%%%%%%%%
\appendices
\section{Eigenvalues of $\tilde{\mathbf{F}}$}
If \bA and \bB are jointly upper- triangularizable, from Schur decomposition, we have
\begin{align}
\mathbf{A}=\mathbf{U}\mathbf{T}_{1}\mathbf{U}^{*},\\
\mathbf{B}=\mathbf{U}\mathbf{T}_{2}\mathbf{U}^{*},
\end{align}
where $\mathbf{U}$ is unitary matrix, and $\mathbf{T}_{1}$ and $\mathbf{T}_{2}$ are upper-triangular matrices.\

Thus, we have
\begin{align}
\tilde{\mathbf{F}}=\mathbf{I}\otimes \mathbf{F}+\left(\mathbf{U}\mathbf{T}_{1}\mathbf{U}^{*}\right)\otimes \mathbf{H}+\left(\mathbf{U}\mathbf{T}_{2}\mathbf{U}^{*}\right)\otimes\mathbf{G}
\end{align}
From properties of Kronecker product we have
\begin{align}
\tilde{\mathbf{F}}=\mathbf{I}\otimes\mathbf{F}&+\left((\mathbf{U}\mathbf{T}_{1})\otimes \mathbf{H}\right)\left(\mathbf{U}^{*}\otimes \mathbf{I}\right)\nonumber\\
&+\left((\mathbf{U}\mathbf{T}_{2})\otimes \mathbf{G}\right)\left(\mathbf{U}^{*}\otimes \mathbf{I}\right)\nonumber\\
=\mathbf{I}\otimes\mathbf{F}&+\left(\mathbf{U}\otimes\mathbf{I}\right)\left(\mathbf{T}_{1}\otimes \mathbf{H}\right)\left(\mathbf{U}^{*}\otimes \mathbf{I}\right)\nonumber\\
&+\left(\mathbf{U}\otimes\mathbf{I}\right)\left(\mathbf{T}_{2}\otimes \mathbf{G}\right)\left(\mathbf{U}^{*}\otimes \mathbf{I}\right)
\end{align}
and we know the eigenvalues of a matrix are the answer to the following equation
\begin{align}
&|\tilde{\mathbf{F}}-s\mathbf{I}|=0,\nonumber\\
&|\left(\mathbf{U}^{*}\otimes \mathbf{I}\right)\left(\mathbf{I}\otimes ( \mathbf{F}-s\mathbf{I})\right)(\mathbf{U}\otimes\mathbf{I})+\mathbf{T}_{1}\otimes \mathbf{H}+\mathbf{T}_{2}\otimes\mathbf{G}|=\nonumber\\
&|\left(\mathbf{U}^{*}\otimes \mathbf{I}\right)\left(\mathbf{U}\otimes( \mathbf{F}-s\mathbf{I})\right)+\mathbf{T}_{1}\otimes \mathbf{H}+\mathbf{T}_{2}\otimes\mathbf{G}|=\nonumber\\
&|\left(\mathbf{U}^{*}\mathbf{U}\right)\otimes \left( \mathbf{F}-s\mathbf{I}\right)+\mathbf{T}_{1}\otimes \mathbf{H}+\mathbf{T}_{2}\otimes\mathbf{G}|=0.
\end{align}
Since $\mathbf{T}_{1}$ and $\mathbf{T}_{2}$ are triangular and $\bU\bU^*=\bI$, hence their diagonal entries are their eigenvalues. Hence the solution for the equation can be broke down to
\begin{align}\label{eq: BlockMat}
|\mathbf{F}+\lambda_{i}\mathbf{H}+\mu_{i}\mathbf{G}-s\mathbf{I}|=0, ~~~i=\{1,~\cdots,~N\},
\end{align}
where $\mu_{i}$'s and $\lambda_{i}$'s are eigenvalues of matrices $\mathbf{A}$ and $\mathbf{B}$, respectively. One of immediate results of (\ref{eq: BlockMat}) is that eigenvalues of $\tilde{\mathbf{F}}$ are the union of eigenvalues of $\mathbf{F}+\lambda_{i}\mathbf{H}+\mu_{i}\mathbf{G}$ for $i=1,~\cdots,~N$.

\bibliographystyle{IEEETran}
\bibliography{mybib}

\begin{thebibliography}{10}
\providecommand{\url}[1]{#1}
\csname url@samestyle\endcsname
\providecommand{\newblock}{\relax}
\providecommand{\bibinfo}[2]{#2}
\providecommand{\BIBentrySTDinterwordspacing}{\spaceskip=0pt\relax}
\providecommand{\BIBentryALTinterwordstretchfactor}{4}
\providecommand{\BIBentryALTinterwordspacing}{\spaceskip=\fontdimen2\font plus
\BIBentryALTinterwordstretchfactor\fontdimen3\font minus
  \fontdimen4\font\relax}
\providecommand{\BIBforeignlanguage}[2]{{%
\expandafter\ifx\csname l@#1\endcsname\relax
\typeout{** WARNING: IEEEtran.bst: No hyphenation pattern has been}%
\typeout{** loaded for the language `#1'. Using the pattern for}%
\typeout{** the default language instead.}%
\else
\language=\csname l@#1\endcsname
\fi
#2}}
\providecommand{\BIBdecl}{\relax}
\BIBdecl

\bibitem{Winfree67}
A.~T. Winfree, ``Biological rhythms and the behavior of populations of coupled
  oscillators,'' \emph{Journal of Theoretical Biology}, vol.~16, no.~1, pp. 15
  -- 42, 1967.

\bibitem{Manaffam14}
S.~Manaffam and A.~Seyedi, ``Probability of stability of synchronization in
  random networks of mismatched oscillators,'' \emph{Arxiv, available:
  http://arxiv.org/abs/1407.7273}, pp. 1--10, 2014.

\bibitem{SyncNet}
\BIBentryALTinterwordspacing
A.~Arenas, A.~Diaz-Guilera, J.~Kurths, Y.~Moreno, and C.~Zhou,
  ``{Synchronization in complex networks},'' \emph{0805.2976}, vol. 469, no.~3,
  pp. 93--153, May 2008. [Online]. Available:
  \url{http://arxiv.org/abs/0805.2976}
\BIBentrySTDinterwordspacing

\bibitem{NavigationSmall}
M.~Fraceschetti and R.~Meester, ``Navigation in small-world networks: a
  scale-free continuum model,'' \emph{Journal of Applied Probability}, vol.~43,
  pp. 1173--1180, 2006.

\bibitem{Manaffam13a}
S.~Manaffam and A.~Seyedi, ``Synchronization probability in large complex
  networks,'' \emph{Circuits and Systems II: Express Briefs, IEEE Transactions
  on}, vol.~60, no.~10, pp. 697--701, Oct 2013.

\bibitem{MasterEq}
L.~M. Pecora and T.~L. Carroll, ``Master stability for synchronized coupled
  system,'' \emph{Physical Review Letters}, vol.~80, no.~10, pp. 2109--2112,
  1998.

\bibitem{weightedNet}
C.~Zhou, A.~E. Motter, and J.~Kurths, ``Universality in the synchronization of
  weighted random networks,'' \emph{Phys. Rev. Lett.}, vol.~96, no.~3, p.
  034101, Jan 2006.

\bibitem{Networks}
M.~E.~J. Newman, \emph{Networks: An Introduction}.\hskip 1em plus 0.5em minus
  0.4em\relax Oxford University Press, 2010.

\bibitem{Heterogeneity1}
T.~Nishikawa, A.~E. Motter, Y.-C. Lai, and F.~C. Hoppensteadt, ``Heterogeneity
  in oscillator networks: Are smaller worlds easier to synchronize?''
  \emph{Phys. Rev. Lett.}, vol.~91, no.~1, p. 014101, Jul 2003.

\bibitem{NatureSmall}
D.~J. Watts and S.~H. Strogatz, ``Collective dynamics of {\bf
  ``small-world''}networks,'' \emph{Letters to Nature}, vol. 393, pp. 440--442,
  1998.

\bibitem{Boundcomplex}
\BIBentryALTinterwordspacing
A.~E. Motter, ``Bounding network spectra for network design,'' \emph{New
  Journal of Physics}, vol.~9, no.~6, p. 182, 2007. [Online]. Available:
  \url{http://stacks.iop.org/1367-2630/9/i=6/a=182}
\BIBentrySTDinterwordspacing

\bibitem{trig2010}
\BIBentryALTinterwordspacing
M.~Mazo, A.~Anta, and P.~Tabuada, ``An iss self-triggered implementation of
  linear controllers,'' \emph{Automatica}, vol.~46, no.~8, pp. 1310 -- 1314,
  2010. [Online]. Available:
  \url{http://www.sciencedirect.com/science/article/B6V21-507665D-3/2/e91c759ba3e4ea6bbf65758a8b13cf04}
\BIBentrySTDinterwordspacing

\bibitem{NCS2011}
X.~Wang and M.~D. Lemmon, ``Event-triggering in distributed networked control
  systems,'' \emph{Automatic Control, IEEE Transactions on}, vol.~56, no.~3,
  pp. 586 --601, 2011.

\bibitem{Manaffam12}
S.~Manaffam, M.~Razeghi-Jahromi, and A.~Seyedi, ``Stabilizing a random dynamics
  network with a random communications network,'' in \emph{Decision and Control
  (CDC), 2012 IEEE 51st Annual Conference on}, Dec 2012, pp. 746--751.

\bibitem{Manaffam13b}
S.~Manaffam and A.~Seyedi, ``Pinning control for complex networks of linearly
  coupled oscillators,'' in \emph{American Control Conference (ACC), 2013},
  June 2013, pp. 6364--6369.

\bibitem{Manaffam13c}
------, ``Pinning control of diffusively coupled oscillators in fast switching
  networks,'' in \emph{Decision and Control (CDC), 2013 IEEE 52nd Annual
  Conference on}, Dec 2013, pp. 6550--6554.

\bibitem{Manaffam15}
S.~Manaffam, A.~Seyedi, A.~Vosoughi, and T.~Javidi, ``Bounded stability in
  networked systems with parameter mismatch and adaptive decentralized
  estimation,'' in \emph{Communication, Control, and Computing (Allerton), 2015
  53nd Annual Allerton Conference on}, Sept 2015.

\end{thebibliography}

\end{document}